\documentclass[twocolumn,prl,showpacs,preprintnumbers,amsmath,amssymb]{revtex4}
\usepackage{graphicx,epsf}% Include figure files
\setkeys{Gin}{width=8.6cm,keepaspectratio}
\usepackage{dcolumn}% Align table columns on decimal point
\usepackage{bm}% bold math

\begin{document}

\title{Magnetic Vacancy Percolation in Dilute Antiferromagnets}

\author{W. C. Barber, F. Ye, and D. P. Belanger}
\affiliation{Department of Physics, University of California,\\
Santa Cruz, California 95064}
\author{J. A. Fernandez-Baca}
\affiliation{Solid State Division, Oak Ridge National Laboratory,\\
Oak Ridge, Tennessee 37831}

\date{\today}
\begin{abstract}
Neutron scattering experiments at the magnetic
vacancy percolation threshold concentration, $x_v$,
using the random-field Ising crystal
$\rm Fe_{0.76}Zn_{0.24}F_2$, show stability of the
transition to long-range order up to
fields $H=6.5$~T.
The observation of the stable long-range order
corroborates the sharp boundary observed in
computer simulations at $x_v$ separating
equilibrium critical scattering behavior
at high magnetic concentration from low concentration
hysteretic behavior.
Low temperature $H>0$ scattering line shapes exhibit
the dependence on the scattering wavevector
expected for percolation threshold fractal structures.
\end{abstract}


\maketitle

The dilute, anisotropic antiferromagnet
$\rm Fe_xZn_{1-x}F_2$
is an extensively studied prototype of the three dimensional ($d=3$)
random-field Ising model (RFIM) \cite{b00}.
As a result of the magnetic vacancies, the magnetic moment is
not uniform and this allows a strong coupling to
to an external magnetic field applied along the spin-ordering
direction.  This constitutes the mechanism for the generation of
random fields \cite{fa79}.  It was shown that such a system
is in the same universality class as a
pure Ising magnet with random fields imposed \cite{c84}.
Settling the question of universality of the phase transition
does not, however, address the effect of
vacancies on microscopic domain formation, which can
mask the phase transition in scattering
experiments.  Such microdomain formation,
which occurs since domain walls can take
advantage energetically of the vacancies,
needs to be well understood in order to
properly interpret the RFIM behavior
of dilute magnets.

For many years controversy surrounded the interpretation of neutron
scattering experiments \cite{b00} on the RFIM critical behavior of dilute
anisotropic antiferromagnets in external magnetic fields, $H$, particularly
$\rm Fe_{x}Zn_{1-x}F_2$ and its less anisotropic isomorph
$\rm Mn_{x}Zn_{1-x}F_2$.  All of these studies, regardless of
whether traditional scaling or various phenomenological
models were used in the interpretations, were done at
concentrations $x \le 0.75$ \cite{brkj85,hfbt93}.  This was natural
since the strength of the random field increases with dilution
and available field strengths required high vacancy concentrations
to readily create suitably strong random fields.  It was, of
course, realized that no ordering would take place for magnetic
concentration below the magnetic percolation threshold
concentration, $x_p=0.246$.  The magnetic {\em vacancy}
percolation threshold concentration occurs at $x_v=1-x_p=0.754$.
Below this concentration, vacancies form a cluster that spans the
crystal.  The
significant role of magnetic vacancy percolation in the formation
of microdomains was not fully appreciated until recently \cite{bb00} and,
prior to that, it was
widely assumed that microdomain formation was an intrinsic
property of the RFIM as realized in
dilute antiferromagnets.
The microdomain structures for small $x$
have been studied extensively \cite{hbkn92,bjkn87}.  It has
recently been shown that these structures
play a crucial role in exchange-bias
structures important to magnetic
recording technology \cite{mgkbgnu00,zdhcc99}.
Microscopic domain structure, for which the characteristic
length scale is small compared to the instrumental
resolution, masks the
neutron scattering critical behavior for two reasons.
First, the scattering contribution from microscopic
domains is superimposed on the scattering from
thermal fluctuations, making it futile to separate
the two.  Second, there is a concomitant decrease in
the Bragg scattering, which consequently no longer represents
the strength of the RFIM order parameter.
This has been particularly frustrating, since
characterization of the RFIM transition is important
in light of the present disagreements between simulations
and experiments \cite{bb00}.

The critical behavior of $\rm Fe_{0.93}Zn_{0.07}F_2$ using
neutron scattering techniques provided evidence \cite{sbf99}
that microscopic domains could be avoided altogether
by doing measurements at high magnetic concentrations,
although very high quality crystals and high fields are
required.  Further experiments have been done
using $\rm Fe_{0.85}Zn_{0.15}F_2$ and $\rm Fe_{0.87}Zn_{0.13}F_2$ \cite{ymkyfb02,yzllbgl02}.
These experiments are
providing the avenue for a complete experimental characterization
of the RFIM universal critical behavior.
It has become quite clear that the behavior at large $x$
is quite distinct from that at low $x$ which exhibits
microdomain structure.
Computer simulations \cite{bb00} were done
to model the behavior of the formation
of microdomains and long-range order in $\rm Fe_{x}Zn_{1-x}F_2$
in an attempt to understand how the behavior crosses from
one type of behavior to the other.
It is suggested by these simulations that low-temperature
metastability and microscopic domain formation
vanish abruptly above $x=0.76$, which closely coincides with the concentration
of the magnetic vacancy percolation threshold concentration,
$x_v=0.754$.  Apparently, the percolating lattice of vacancies
results in the instability of long-range order
below the transition.  In previous experiments, little attention has
been paid to the percolation
of magnetic vacancies.
In light of its importance to the understanding of
the RFIM we were motivated to investigate the
scattering in $\rm Fe_{0.76}Zn_{0.24}F_2$, which is very
close to the critical concentration $x_v$.  The concentration
was determined using density measurements and the concentration
gradient of a few tenths of a percent was determined using room
temperature birefringence techniques \cite{kfjb88}.

Considerable focus has been given to
the study of behavior near the complementary
threshold concentration for magnetic percolation,
$x_p=1-x_v$, using neutron scattering, specific heat,
linear birefringence, magnetization
and ac susceptibility
techniques \cite{by93,mcr89,brc00,bb00b,jdnb97}.  The
value $x_p=0.246$ is based on a calculation
including only the dominant $J_2$ interaction \cite{se64}.
However, for concentrations close to
$x_p$, the system is extremely sensitive
to very weak interactions that are 
insignificant away from $x_p$.  Spin-glass-like behavior at
$H=0$ has been well characterized.
Even far above $x_p$, large fields
cause a crossover from the low-field
microdomain-dominated random-field behavior
to the spin-glass-like behavior
\cite{mkjhb91,bmmkje91,mltl99,mltl98,rfm00,satk88}.

The neutron scattering experiments were performed at the Oak Ridge
National Laboratory High Flux Isotope Reactor using a double-axis
spectrometer configuration.  The beam was horizontally collimated
to $20$ min of arc before and after the sample and
48 min of arc before the monochromator.  The neutron
energy was either $13.7$~meV on the HB1 spectrometer or $14.7$~meV on
the HB1A spectrometer.  Higher energy neutrons
were eliminated using pyrolytic graphite filters.  Most of the data were taken
with transverse scans about the $(1 0 0)$ antiferromagnetic Bragg point.
The $\rm Fe_{0.76}Zn_{0.24}F_2$ crystal
has an irregular shape approximately $4 \times 5 \times 10$~mm.
It has a resolution limited Bragg
peak, but very small secondary peaks appear for $q >0$ in the
low temperature scans.  Near the transition, these tiny peaks are
not evident.  All the data used in the analysis of the line
shapes at low temperatures are on the $q<0$ side of the Bragg peak,
where no hint of any secondary peaks are observed. 
The thermometry was based on a commercially calibrated carbon
thermometer.  Two primary thermal cycling procedures
that are often employed to investigate hysteresis in
the RFIM include: 1) cooling in the absence of a field, raising the
field and warming through the transitions (ZFC); and 2)
cooling in the field (FC).

Figure 1 shows scattering intensity vs.\ q, in reciprocal
lattice units (rlu), at $H=3$ and $5$~T close to the
transition temperatures $T_c=61.6$ and $60.4$~K, respectively.
The transition at $H=0$ is at $63.2$~K.
Whereas the critical scattering from samples with
$x<x_v$ exhibits strong hysteresis,
it is clear that the $|q|>0$ critical scattering shown in
Fig.\ 1 is free from hysteresis and that indicates
there is no microscopic domain structure frozen in upon FC.

For experiments free of extinction effects, the
magnetic Bragg scattering intensity is expected to follow
the power law behavior
\begin{equation}
I = {M_s}^2 \sim |t|^{2\beta}
\end{equation}
where $M_s$ is the staggered magnetization and
$\beta \approx 0.35$ for the random-exchange model
and $\beta \approx 0.16$ the RFIM \cite{yzllbgl02}.
However, neutron scattering in high-quality bulk crystals
can suffer from severe extinction; the beam is
depleted of neutrons that satisfy the Bragg condition
and the scattering intensity is therefore saturated
and cannot exhibit the correct $T$ dependence.  The
extinction effects usually preclude determination of a reliable value
of $\beta$, the critical exponent for the staggered magnetization,
from an analysis of the neutron scattering
data in very high quality bulk crystals.  The Bragg scattering
does show hysteresis, which indicates incomplete
FC ordering on very long length scales, relative to
the instrumental resolution.  Such hysteresis occurs for all $x$
and is likely a consequence of the slow activated dynamics
\cite{f86,kmj86,nkj91,bkk95}
of the RFIM very close to $T_c(H)$.

\begin{figure}[t]
\centerline{
\epsfxsize=9cm %width
\epsfbox{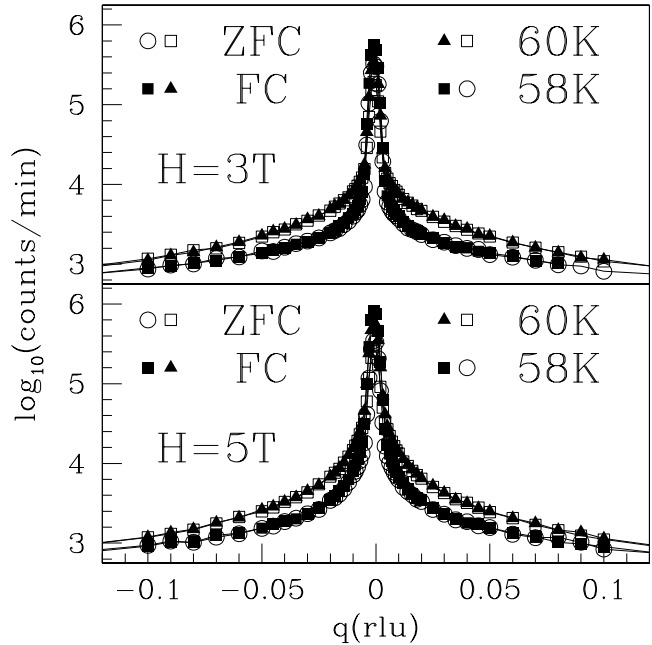}
}
\caption{ZFC and FC neutron scattering at $H=3$ and $5~T$
below but close to the transition temperature.
The contributions include a constant background, critical
scattering which is seen to increase as the temperature
approaches $(T_c(H))$, and a Bragg peak due to long range order.
}
\end{figure}

The ZFC Bragg intensity, corrected for the background determined at
high $T$, is shown in Fig.\ 2 vs.\ $T$ for $H=3$~T.  The inset
of Fig.\ 2 shows the critical scattering at small $q$, but well outside
the transverse instrumental resolution where the
Bragg intensity is negligible.  The scattering line shapes in
this sample are complicated by admixture of critical scattering
and contributions from the vacancy lattice because of
the proximity of this sample to $x_v$, as will be discussed.
Hence, we could not confidently analyze in detail the
critical scattering line shape.  Nevertheless, we obtained an
approximate accounting of the critical scattering intensity by
taking a squared-Lorentzian line shape folded with the appropriate
resolution correction.  An overall amplitude had to be chosen for the
fit shown in the inset.  This same amplitude was applied to the
$q=0$ case and the resulting curve is shown in the main part of Fig.\ 2
as the solid curves.
Subtracting this from the raw data (open symbols) yields the corrected
Bragg scattering data (filled symbols).  Taking into account
the concentration gradient rounding of a few tenths of a percent,
it is quite apparent that
the Bragg intensity data approach $T_c(H)$
with a steep slope, in contrast with
scattering experiments\cite{bwshnlrl96} with $x<x_v$, where
the slope is nearly zero.
Although we cannot analyze the data according to Eq.\ 1 to
obtain $\beta$ because of extinction, we may still
conclude from the shape of the Bragg intensity vs.\ T that
this sample does not form microscopic domain structure.
This is consistent with the lack of hysteresis in the critical
scattering shown in Fig.\ 1.

\begin{figure}
\centerline{
\epsfxsize=9cm %width
\epsfbox{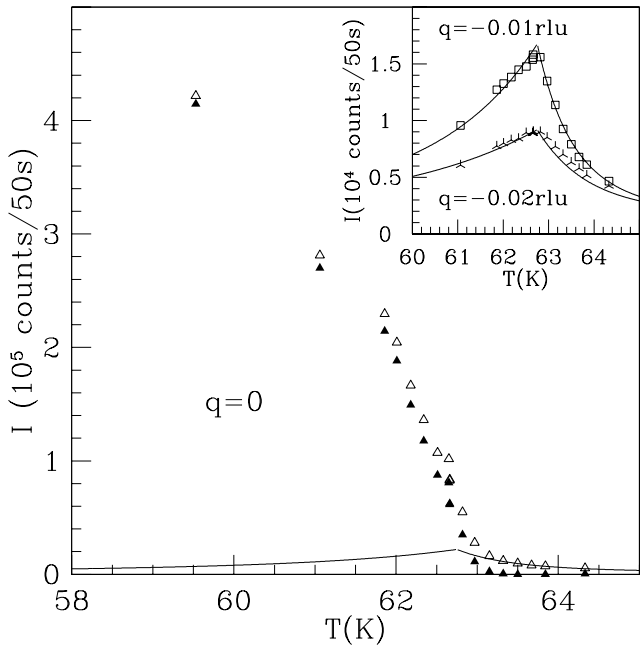}
}
\caption{The ZFC antiferromagnetic Bragg peak for $H=3~T$.
The contribution at $q=0$ due to critical fluctuations is subtracted
from the data represented by the open symbols to give the
corrected intensities represented by the solid symbols. 
}
\end{figure}

At the percolation threshold concentrations, magnetic sites or
magnetic vacancy sites form fractal structures.  In either case,
scattering from the fractal structure will exhibit a power law
behavior \cite{sa94,iifn95}
\begin{equation} 
I_f \sim q^{-2.53} \quad .
\end{equation}
The only
difference is that in the case of magnetic vacancies there is
also a Bragg scattering peak from the average $M_s$.
With magnetic site percolation, the
average $M_s$ is zero at the threshold.  Since we believe the $x=0.76$
crystal is close to $x_v$, we plotted the logarithm of the
scattering intensity vs.\ the logarithm of $-q$ for $q<0$ in
Fig.\ 3.  Only some of the scans are shown; data were taken for
$H=0$ and for $2T<H<7$~T in steps of $0.5$~T. Several interesting
features are evident.  

\begin{figure}
\centerline{
\epsfxsize=9cm %width
\epsfbox{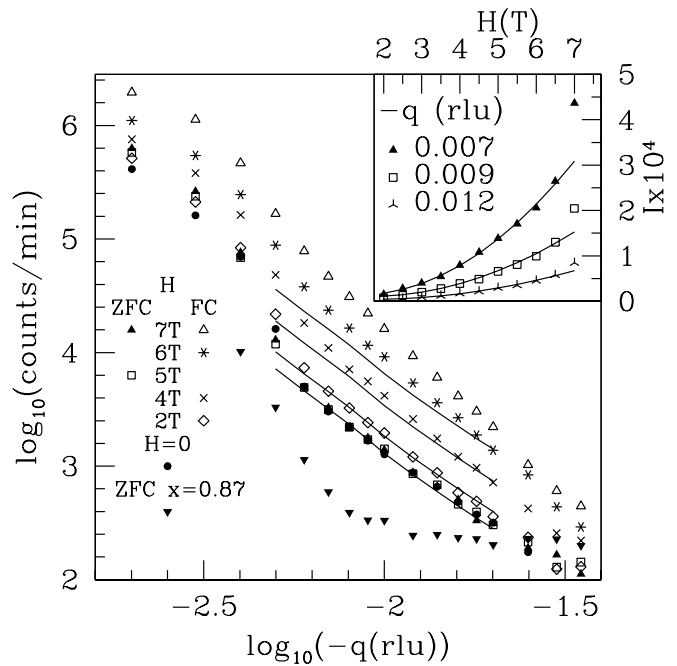}
}
\caption{The logarithm of the neutron scattering intensity vs.\
the logarithm of $q$ for $H=0~T$ and for ZFC and FC up to $H=7~T$
at $T=20K$, well below $T_c(H)$.  The background determined
at large $q$ has been subtracted.  Only a few sets of data are
shown for clarity.  All ZFC data lie on the same curve, but only
$H=5$ and $7$~T data are shown.
}
\end{figure}

As shown explicitly for two fields, $H=5$ and $7$~T in Fig.\ 3,
the ZFC line shapes,
for $T=29$~K or less,
are all identical with the $H=0$ line
shape and are the lowest in intensity.  For comparison, a line
with a fractal exponent of 2.53 for three dimension, with
the spectrometer resolution folded in, is plotted in
the graph with the amplitude adjusted to fit the ZFC intensity data.
It is clear that the ZFC scattering line shapes for
$x=0.76$ follow Eq.\ 2  quite well.  To contrast this behavior, we show
in Fig.\ 3 similar data for a sample with concentration of
$x \approx 0.87$ \cite{ymkyfb02}, indicated by the solid triangles,
for $T=58.5$~K, only $3.4$~K below the transition.  The scattering for
this concentration, well above $x_v$, shows little evidence
of scattering outside the Bragg region as expected since
the vacancies do not form large fractal structures at this
concentration.  The behavior for ZFC shown in Fig.\ 3
suggests strongly that, for $x=0.76$, the scattering is indeed from the
vacancy percolation fractal structure under the ZFC procedure.

The FC data for $x=0.76$ increase in intensity with the applied field.
We compare the data to Eq.\ 2 by adjusting
the amplitude to fit the data at $q=10^{-1.7}$~rlu.  It
is quite clear that for $H>0$ the line shapes
deviate strongly from the behavior in Eq.\ 2, more so as the field
increases.  The inset of Fig.\ 3 shows the deviations of
the intensities from
the curves representing Eq.\ 2 at $q = -0.007$, $-0.009$,
and $-0.012$~rlu as a function of the applied field.
The deviations for $H \le 6.5$~T increase smoothly
with the field.  Two possible sources exist for the
excess scattering.  One is the relief of
extinction.  This has been observed for neutron Bragg scattering in
the RFIM experiments on bulk crystals \cite{b00},
but not for scattering outside the Bragg region.
The other possibility, perhaps more significant, is the
scattering from domains, which coexist with
antiferromagnetic long-range order, that increase in number
with increasing applied field.
Since the total scattering in FC comes from sources in
addition to that of vacancy sites, 
it is difficult to analyze in detail.  

The difference between FC intensities and Eq.\ 2
at $H=7$~T deviate strongly from the
smooth curves describing the data for $H \le 6.5$~T, as shown
in the inset of Fig.\ 3.  This qualitatively new behavior
most likely represents a breakdown of the antiferromagnetic
long-range structure
for $H>6.5$~T.  In such a case many more domains
are introduced into the system, resulting in much more
scattering intensity.
This field is consistent with the increasing field at which
spin-glass-like behavior appears for samples with
$x<x_v$, as shown in Fig.\ 4.
Apparently, even at low temperatures, long-range antiferromagnetic order
for $x=0.76$ is stable upon FC for $H<7$~T, in contrast to the behavior
for $x<x_v$, where metastable domains dominate the scattering
under the FC procedure.  For $x \approx 0.87$,
only a few percent above $x_v$, magnetization experiments \cite{samb02}
indicate that the transition to long-range antiferromagnetic
order breaks down only for fields above $H=18$~T, nearly three
times the field that causes a breakdown in the behavior of the
sample with $x=0.76$, demonstrating the stability of antiferromagnetic
order for $x>x_v$.

\begin{figure}[t]
\centerline{
\epsfxsize=9cm %width
\epsfbox{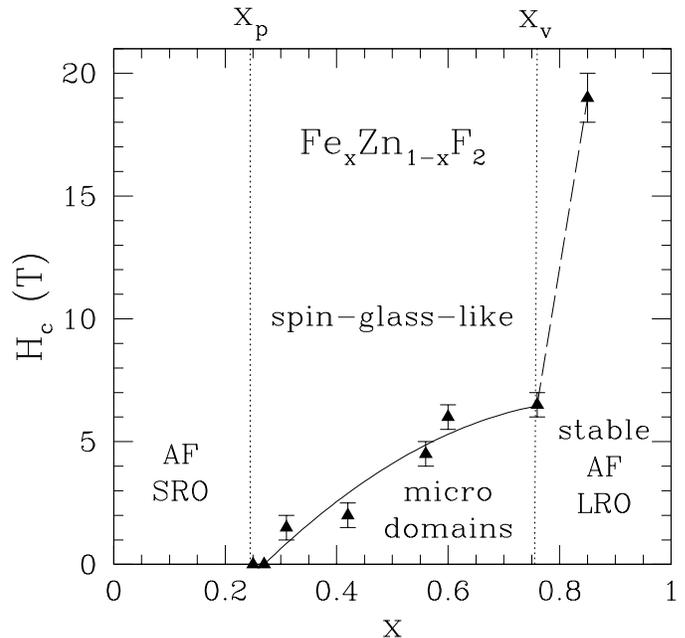}
}
\caption{The concentration dependence of the phases observed
in $\rm Fe_{x}Zn_{1-x}F_2$.  For $x<x_p$, the magnetic
percolation threshold concentration, only antiferromagnetic
short-range order
is possible.  For $x_p<x<x_v$, microscopic domain formation
occurs at low fields and spin-glass-like behavior occurs
at high fields.  For $x>x_v$, antiferromagnetic
long-range order is stable, without microscopic domain formation,
to very large applied fields.  Critical behavior measurements
can only be reliably done for $x>x_v$.  Data are taken from
various experiments cited in the text.
}
\end{figure}

We have shown that the magnetic concentration range for which
equilibrium random field critical scattering is observed for
dilute antiferromagnets in an external field has a lower bound at
$x_v$.  Whereas our $x=0.76$ crystal shows no critical scattering
hysteresis, it is quite evident for slightly smaller magnetic
concentrations \cite{hfbt93}.
Our results suggest that the percolation of
vacancy sites occurring for $x<x_v$ precipitates the formation of
domains below $T_c(H)$ in ZFC preparation as well as in FC,
corroborating the conclusions drawn
from simulations \cite{bb00}.  A theoretical connection between the one
dimensional fractal geometry of vacancy sites at percolation and
three dimensional domains has not been adequately explained from
a theoretical perspective.

It is now evident that
there are three magnetic concentration regimes in $d=3$ dilute
antiferromagnets separated by the percolation threshold
concentrations $x_p$ and $x_v$.  For $x>x_v$, long-range
antiferromagnetic order is stable up to very large magnetic
fields.  For $x_p<x<x_v$ the system is unstable.  At low fields,
the formation of microdomain structure
takes place upon FC for all $T<T_c(H)$ and upon ZFC close to $T_c(H)$.
A spin-glass-like phase forms at high field.  Below $x_p$, there
can be no long-range magnetic order.  It is
the geometry of the lattice in question which defines the
location of these boundaries, and although we study one
particular magnetic lattice type, the body centered tetragonal
structure of $\rm Fe_{x}Zn_{1-x}F_2$, our
results should apply more generally to dilute magnets in an
applied field.

Interestingly, the specific heat behavior is not dependent in an
obvious way on the concentration.  Similar hysteresis upon FC and
ZFC is observed \cite{b00} very close to $T_c(H)$
for concentrations above and below $x_v$.  No specific
heat hysteresis is observed at low $T$.  The
contrast between the relative insensitivity
of the specific heat techniques
with the extreme sensitivity of the scattering techniques is
certainly due to the greater dependence of the scattering on long
length correlations that are greatly affected by domain
formation.  The hysteresis in the case of specific heat is
related to the activated dynamics very close to $T_c(H)$ that
affects the behavior at
all $x$ and not domain formation, which only occurs for $x<x_v$.
 
From the results of this investigation, we conclude that studies
of the random-field phase transition should be conducted with
magnetic concentrations greater than $x_v$.  It is advantageous
to use concentrations not too much greater than this to maximize
the random-fields for available applied fields.  However, if the
concentration is too close to $x_v$, one must take into account
scattering from the magnetic vacancy percolation cluster.  Recent
experiments at $x=0.87$ indicate that, at this concentration,
such scattering is negligible \cite{ymkyfb02,yzllbgl02}.

This work was funded by Department of Energy Grant No.\
DE-FG03-87ER45324 and by the Oak Ridge National Laboratory,
which is managed by UT-Battelle, LLC, for the U.S. Dept.\
of Energy under contract DE-AC05-00OR22725.

\end{document}